\documentclass[12pt]{amsart}
\usepackage[foot]{amsaddr}

\usepackage{graphicx}
\usepackage{amsmath,amsthm,amssymb}
\usepackage{caption}
\usepackage[labelformat=simple]{subcaption}
\usepackage{multirow}
\usepackage{makecell}
\usepackage{color}
\usepackage[left=25.4mm,right=25.4mm,top=30mm,bottom=25.4mm,a4paper]{geometry}

\newcommand{\gen}{\mathrm{gen}}

\newcommand{\lin}{\mathrm{lin}}

\def\alphaBar{\langle \alpha \rangle_T}
\def\rhoBar{\langle \rho \rangle_x}
\def\kappaBar{\langle \kappa \rangle_x}

\def\alphaBar{\overline \alpha }
\def\rhoBar{\overline \rho}
\def\kappaBar{\overline \kappa}
\def\Qh{Q_h}
\def\Qc{Q_c}
\def\QhBar{\overline \Qh }
\def\QcBar{\overline \Qc }

\begin{document}


\title{General Efficiency Theory of Thermoelectric Conversion}


\author{Byungki Ryu}
\email[]{byungkiryu@keri.re.kr}

\author{Jaywan Chung}
\email[]{jchung@keri.re.kr}

\author{SuDong Park}

\address[B Ryu, J Chung, \and SD Park  ]{Energy Conversion Research Center, Korea Electrotechnology Research Institute (KERI), 12, Bulmosanro 10beon-gil, Seongsan-gu, Changwon, 51543, Republic of Korea}
\address[$^{\dagger}$]{B.R. and J.C. contributed equally to this work}


\date{\today}

\begin{abstract}
In this Letter, we show thermoelectric conversion efficiency is \emph{exactly} determined by \emph{three} independent material parameters $Z_\gen$, $\tau$, and $\beta$.
Each parameter is a figure of merit hence improving $\tau$ or $\beta$ is an additional way to increase the efficiency.
The $Z_\gen$ generalizes the traditional figure of merit \emph{zT}. Two degrees of freedom $\tau $ and $\beta$ reflect the temperature gradients of the material properties and are crucial to evaluate the heat current altered by non-zero 
Thomson heat and asymmetric Joule heat escape.
Physical insights on high $\tau$ or $\beta$ materials explain why the single parameter approaches can be inaccurate for efficiency prediction. 
\end{abstract}


\maketitle

A thermoelectric module utilizing the thermoelectric effect for direct conversion of thermal energy into electrical energy is a heat engine \cite{ioffe1957semiconductor}.
Hence its conversion efficiency is the fraction of input heat current $Q_h$ into the electrical power $P$ delivered to an external load. 
The power $P$ is simply determined by the electrical \emph{device parameters} as $P = I(V - I R)$ where $I$, $V$, and $R$ are electric current, total generated voltage, and internal electrical resistance.
However the heat current has no such expression due to the nonlinearity of the thermoelectric equation \cite{chung2014nonlocal, borrego1961optimum, sherman1960calculation, goupil2015continuum} caused by temperature-dependent thermoelectric material properties.
As a result, the thermoelectric efficiency $\eta = \frac{P}{Q_h}$ has no analytical expression in terms of the device parameters.

When the material properties are temperature-independent, the single parameter $zT$ suggested by Ioffe \cite{ioffe1957semiconductor, borrego1961optimum} is a figure of merit for the thermoelectric efficiency.
As its generalization for temperature-dependent material properties, several single average parameters have been suggested \cite{goupil2015continuum, borrego1961optimum, borrego1963carrier, borrego1964approximate, sherman1960calculation, kim2015relationship} but their efficiency prediction can be significantly different from exact numerical one for some practical material curves 
\cite{goupil2015continuum, borrego1961optimum, borrego1963carrier, borrego1964approximate, sherman1960calculation, sunderland1964influence, kim2015relationship, kim2015efficiency, armstrong2016estimating, wee2011analysis}.

In this Letter, we show the input heat current $Q_h$ of a one-dimensional thermoelectric generator model with temperature- and position-dependent material properties is determined by the device parameters and \emph{two additional material parameters} $\tau$ and $\beta$.
As a consequence, the efficiency $\eta$ is determined by three material parameters $Z_\gen$, $\tau$, and $\beta$. Here the $Z_\gen$ is a generalization of the figure of merit $zT$ and is written by the device parameters; see \eqref{genPF}. Although the three material parameters depend on the electrical current $I$, the dependence is negligible when the maximum power or maximum efficiency is considered; see Fig. \ref{parameter-variation}. Treating the $Z_\gen$, $\tau$, and $\beta$ as constants, we have an analytical formula for the maximum thermoelectric efficiency as a natural generalization of the constant material property case; see \eqref{etaMaxGen}. Furthermore, each of our $\tau$ and $\beta$ is a dimensionless figure of merit, hence its improvement is a novel approach to enhance the efficiency, different from improving the $zT$. While the definitions of $\tau$ and $\beta$ in \eqref{tau-beta} are involved, their approximations \eqref{one-shot-approx} give clear insights to improve the $\tau$ and $\beta$, and explain why the single parameter generalizations of $zT$ fail for some materials, as discussed later.

Before deriving the $\tau$ and $\beta$, first we observe the necessity of them. Let $\alpha$, $\rho$, $\kappa$ be temperature-dependent Seebeck coefficient, electrical resistivity, and thermal conductivity of a thermoelectric material, respectively. For a single thermoelectric leg module with length $L$ and cross-sectional area $A$ in one spatial dimension, we may define \emph{average parameters} of the material properties as 
$\alphaBar := \frac{ V }{\Delta T} = - \int \alpha \frac{dT}{dx} \,dx$,
$\rhoBar := \frac{A}{L} R = \frac{1}{L} \int \rho \,dx$,
and $\frac{1}{\kappaBar} := \frac{A}{L} \frac{1}{K} = \frac{1}{L} \int \frac{1}{\kappa} \,dx$.
Here $\Delta T$ is the given temperature difference between the two ends of the generator, $x$ is the spatial coordinate inside the module, $T=T(x)$ is the temperature distribution inside the module, and $1/K$ is the thermal resistance of the module. If the material properties do not depend on $T$, the hot-side input heat current at temperature $T_h$ and the cold-side output heat current at temperature $T_c$ are determined by the average parameters as
$\QhBar =  K \Delta T + I \alphaBar T_h  - \frac{1}{2} I^2 R $ and
$\QcBar =  K \Delta T + I \alphaBar T_c  + \frac{1}{2} I^2 R $.
The electrical power delivered outside the module is $P =\QhBar - \QcBar $. If the material properties \emph{depend on $T$} while the average properties remain unchanged, the heat currents change to different values $Q_h$ and $Q_c$ but the power, which is determined by the average properties, does not change. This implies $P = Q_h - Q_c = \QhBar - \QcBar$.
Hence there exists a \emph{backward heat current $B$} such that $Q_h = \QhBar -B$ and $Q_c = \QcBar - B$.
Meanwhile, since there are two heat sources in the thermoelectric equation, i.e. Thomson heat porportional to electrical current $I$ and Joule heat proportional to $I^2$, the $B$ should be decomposed into $I$ and $I^2$ terms which reflect the effective Thomson heat flow and asymmetric Joule heat escape.
Our $\tau$ and $\beta$ in \eqref{tau-beta} are these contribution terms satisfying
$B = \left(I \alphaBar  \Delta T \right) \tau   + \left( \frac{1}{2} I^2 R \right) \beta$ and
\begin{equation}\label{QhQc}
\begin{aligned}
Q_h &= K \Delta T + I \alphaBar ( T_h - \tau \Delta T )    - \frac{1}{2} I^2 R ( 1+\beta ), \\
Q_c &= K \Delta T + I \alphaBar ( T_c - \tau \Delta T )  + \frac{1}{2} I^2 R ( 1-\beta ).
\end{aligned}
\end{equation}

To derive the $\tau$ and $\beta$, we examine the thermoelectric equation. The thermoelectric effect is expressed in terms of electric current density $J := I/A$ and heat current density $J^Q := Q / A$ as $J = \sigma ( E - \alpha \nabla T)$ and $J^Q =  \alpha T J - \kappa \nabla T$ where $E$ is the electric field. 
Applying the charge and energy conservation laws to $J$ and $J^Q$,
we can obtain the thermoelectric differential equation of temperature $T$ in a one-dimensional thermoelectric leg without radiative and convective losses \cite{chung2014nonlocal, goupil2015continuum}:
\begin{equation}\label{TEQ-PDE-1D}
\frac{d}{dx} \left(\kappa \frac{dT}{dx} \right) -T \frac{d \alpha}{dT} \frac{dT}{dx} J +  \rho J^2  = 0,
\end{equation}
where $x$ is the spatial coordinate inside the leg. The left-hand side of \eqref{TEQ-PDE-1D} is composed of thermal diffusion, Thomson heat generation, and Joule heat generation.
Let $f_T$ be the heat source term of \eqref{TEQ-PDE-1D} as $f_T (x) :=-T \frac{d \alpha}{dT} \frac{dT}{dx} J +  \rho J^2 $.
Assuming the $\kappa(x)$ and $f_T(x)$ are known, the linear equation \eqref{TEQ-PDE-1D} with a Dirichlet bounday condition $T(0)=T_h$, $T(L)=T_c$ can be solved to find an \emph{integral equation} for $T$ and $\frac{dT}{dx}$:
\begin{equation}\label{TgradT}
\begin{aligned}
T(x) &=  \left( T_h - \frac{K \Delta T}{A} \int_0^x \frac{1}{\kappa (s)} \,ds  \right)\\
&\quad +  \left( -\int_0^x \frac{F(s)}{\kappa(s)}\,dx +\frac{ K \,\delta T}{A} \int_0^x \frac{1}{\kappa (s)} \,ds \right ) ,\\
\frac{dT}{dx}(x) &=  \left(  -\frac{K \Delta T}{A}  \frac{1}{\kappa (x)} \right)
+  \left( \frac{F(x)}{\kappa(x)} +\frac{ K \,\delta T}{A} \frac{1}{\kappa (x)}  \right ) ,
\end{aligned}
\end{equation}
where $F(x) := \int_0^x \ f_T (s)  \,ds$ and $\delta T :=  \int_0^L \frac{F(x)}{\kappa (x)} \,dx$.
Note that the first equation is of the form $T = \varphi [T]$ where $\varphi$ is an integral operator. With this relation, the exact $T$ can be obtained via fixed-point iteration $T_{n+1} = \varphi [T_{n}] $ \cite{burden2010numerical} in most practical cases.

Using the second equation in \eqref{TgradT} for $\frac{dT}{dx}(x)$, we derive an integral equation of the heat currents as
\begin{equation} \label{SI-qHot}
\begin{aligned}
Q_h &= I \alpha_h T_h - A \kappa_h \Big(\frac{dT}{dx}\Big)_h = I \alpha_h T_h + K (\Delta T - \delta T), \\
Q_c &= I \alpha_c T_c - A \kappa_c \Big(\frac{dT}{dx}\Big)_c = Q_h  - P .
\end{aligned}
\end{equation}
The $\delta T$ has two contribution terms of $I$ and $I^2$ from double integration of Thomson and Joule heat: since
\[\begin{split}
F_T(x) &=  I^2 \int_0^x \frac{1}{A^2} \rho(s)\,ds - I \int_0^x \frac{1}{A} T(s) \frac{d \alpha}{dT}(T(s)) \frac{dT}{dx}(s) \,ds \\
&=: I^2 F_T^{(2)}(x) - I F_T^{(1)}(x),
\end{split}\]
we have
\[\begin{split}
\delta T &= I^2 \int_0^L \frac{F_T^{(2)}(s)}{\kappa(s)}\,ds - I \int_0^L \frac{F_T^{(1)}(s)}{\kappa(s)}\,ds \\
&=: I^2 \delta T^{(2)} - I \delta T^{(1)}.
\end{split}\]
Rewriting the $Q_h$ in \eqref{SI-qHot} into the form of \eqref{QhQc}, we obtain
\begin{equation}
\begin{aligned}  \label{tau-beta}
\tau &:= \frac{1}{\alphaBar \Delta T} \left[ (\alphaBar - \alpha_h) T_h -K \,\delta T^{(1)} \right], \\
\beta &:= \frac{2}{R} K \,\delta T^{(2)} - 1.
\end{aligned}
\end{equation}
For $T$-independent material properties, $\delta T^{(2)} = \frac{1}{2}\frac{R}{K}$ and $\delta T^{(1)} \equiv 0$ so that $\tau \equiv 0$ and $\beta \equiv 0$, which implies $Q_h = \QhBar = K \Delta T + I \alphaBar T_h -\frac{1}{2} I^2 R$, as expected. 

Next we define the $Z_\gen$ as a generalization of the figure of merit $zT$.
For a given load resistance $R_{\rm L}$, the electric current is $I = \frac{V}{ R \left( 1+ \gamma \right) }$ where $\gamma := R_{\rm L}/R$.
Hence the power delivered to the load is $P = I^2 R_{\rm L} = I ( V - I R ) = \frac{ \alphaBar^2 }{\rhoBar} \frac{(\Delta T)^2 }{L/A} \frac{\gamma}{\left( 1+\gamma \right)^2 }$. With this observation, we define the \emph{general device power factor} as ${P\!F}_{\rm \!gen} := \frac{\alphaBar^2}{\rhoBar}$ and the \emph{general device figure of merit} as 
\begin{equation}\label{genPF}
Z_\gen := \frac{(V/\Delta T)^2}{RK} = \frac{\alphaBar^2}{\rhoBar \,\kappaBar}.
\end{equation}
When the ${P\!F}_{\rm \!gen}$ is slowly varying on $I$, the power $P$ is maximized near $\gamma = 1$.
The $Z_\gen$ is adopted for the simplication of the efficiency formula as in \eqref{eta-formula}.
Furthermore, the $Z_\gen$ is a generalization of the previous single average parameters.
The $Z_\gen$ is reduced to the average figure of merit $z_{\mathrm{av}} := \frac{\left<\alpha\right>^2}{\left<\rho\kappa\right>} $ by Ioffe and Borrego \cite{ioffe1957semiconductor, sherman1960calculation, borrego1961optimum, borrego1963carrier, borrego1964approximate} under the constant heat approximation. Here the bracket $\left<\cdot\right>$ indicates the averaging over $T$.
The $Z_\gen$ is reduced to the engineering figure of merit $ Z_{\rm eng}:= \frac{\left<\alpha\right>^2}{\left<\rho \right> \left< \kappa\right> }$ by Kim \emph{et al.} \cite{kim2015relationship} under the linear-$T$ approximation. Similarly, the ${P\!F}_{\rm \!gen}$ is reduced to the effective power factor 
$ \frac{\left<\alpha\right>^2 \left< \kappa  \right>}{\left<\rho\kappa\right>}$ under the constant heat approximation\cite{zabrocki2012effective} .

\begin{figure}
\centering\includegraphics[width=0.8\textwidth]{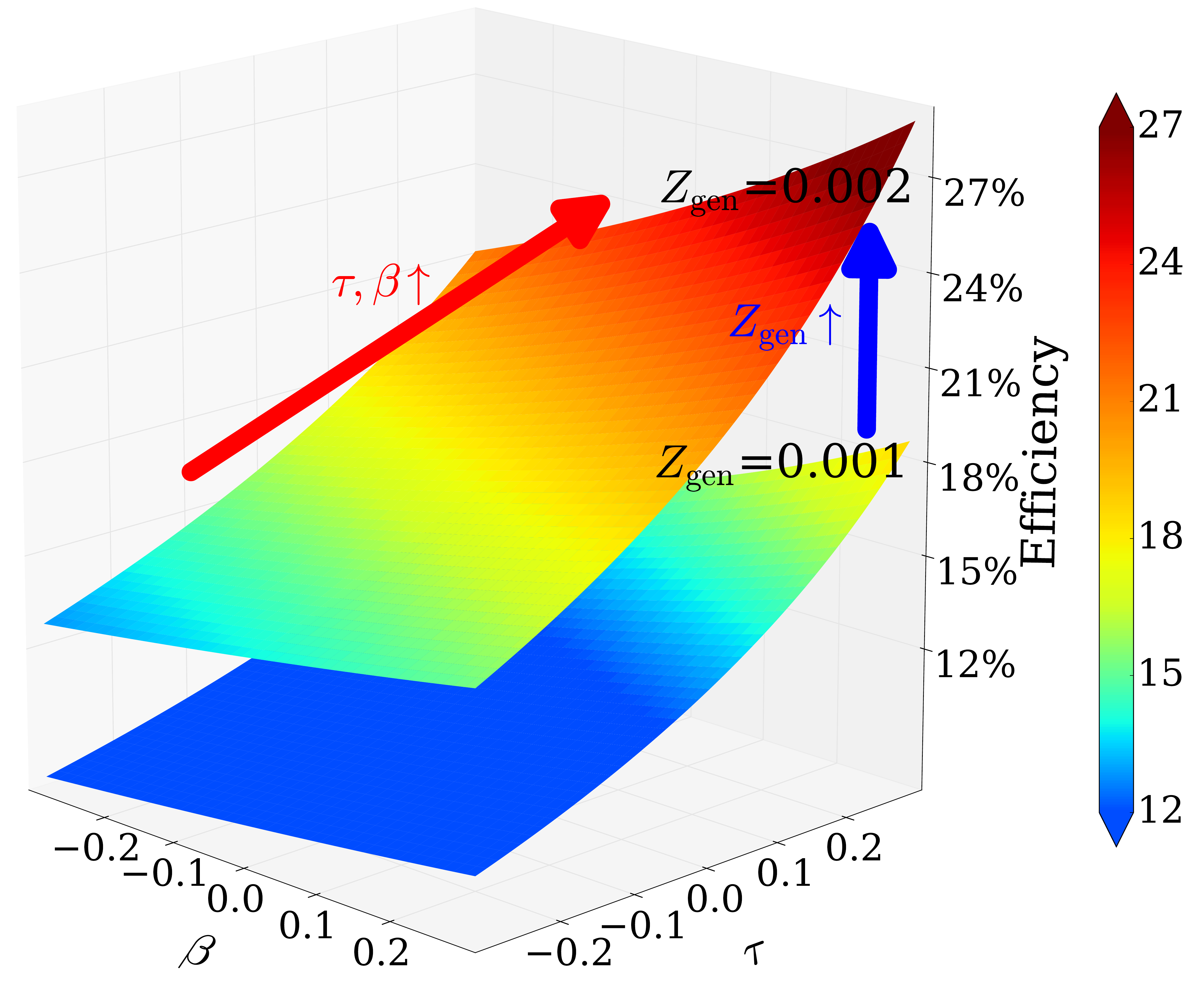}
\caption{Maximum thermoelectric efficiency surface in equation \eqref{etaMaxGen} is drawn for $T_h = 900 K$ and $T_c = 300 K$, with $Z_\gen = 0.002 K^{-1}$ and $0.001 K^{-1}$.
Improving one of $Z_\gen$, $\tau$, and $\beta$ increases the efficiency.}
\label{FIG1}
\end{figure}

\begin{figure}
\centering\includegraphics[width=0.8\textwidth]{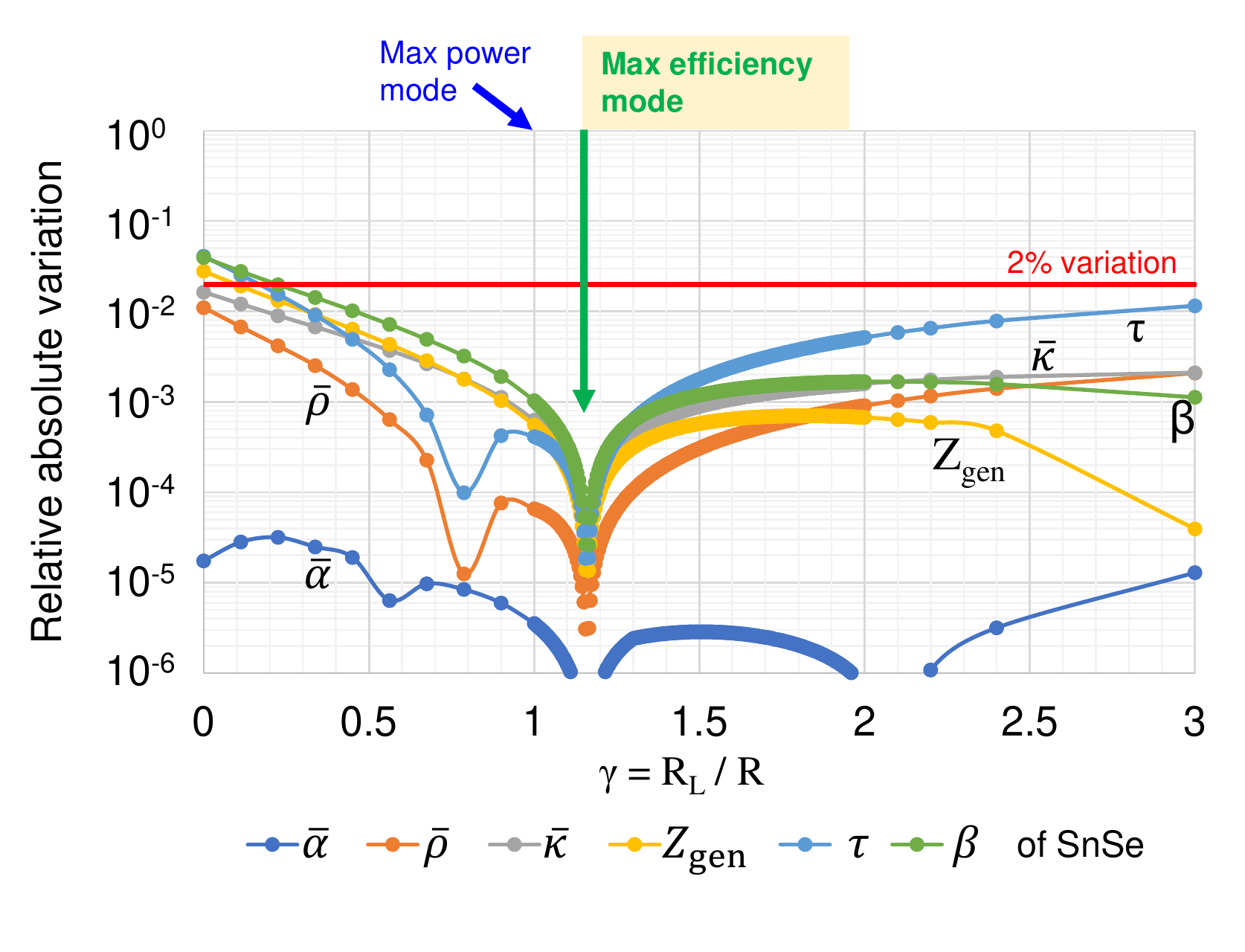}
\caption{Relative absolute variations of three average parameters ($\alphaBar$, $\rhoBar$, $\kappaBar$) and three thermoelectric degrees of freedom ($Z_\gen$, $\tau$, $\beta$) are calculated with respect to the resistance ratio $\gamma = R_L / R$ for a single-element leg generator module using {\rm SnSe} material properties \cite{zhao2014ultralow}.}
\label{parameter-variation}
\end{figure}

Finally we consider the thermoelectric conversion efficiency $\eta = \frac{Q_h - Q_c }{Q_h}$. 
Using \eqref{QhQc}, we can verify that the dimensionless heat currents $Q_h / K \Delta T $, $Q_c / K \Delta T $, and the efficiency are
determined by five parameters $\Delta T$, $Z_\gen$, $\tau$, $\beta$, and $I$ (or the resistance ratio $\gamma$).
Thereby, given external thermal and electrical condition, the efficiency is \emph{exactly} determined by three parameters as
\begin{equation}\label{eta-formula}
\begin{split}
&\eta (Z_\gen, \tau, \beta | T_h, T_c, \gamma) \\
&= \frac{  \Delta T  \frac{\gamma}{(1+ \gamma)^2 } }{
 1/Z_\gen + \big( \frac{1}{1+ \gamma} \big) ( T_h - \tau \Delta T ) 
 -\frac{1}{2} \Delta T \big( \frac{1}{1+ \gamma} \big)^2 (1+\beta).} 
\end{split}
\end{equation}
Note that the $\eta$ is monotonically increasing with respect to $\Delta T$, $Z_\gen$, $\tau$ and $\beta$. The monotone increasing property on $\tau$ and $\beta$ is easily observed from $Q_h$ in \eqref{QhQc}; since $Q_h$ is monotonically decreasing with respect to $\tau$ and $\beta$, the $\eta=\frac{P}{Q_h}$ is monotonically increasing.
Therefore each of $Z_\gen$, $\tau$, and $\beta$ is a figure of merit; see Fig. \ref{FIG1}.

The efficiency formula \eqref{eta-formula} is applicable to segmented- and graded-material devices with contact resistance. This is because the computation of $Z_\gen$, $\tau$, $\beta$ in \eqref{genPF} and \eqref{tau-beta} is based on the \emph{integral formulation} of temperature distribution in  \eqref{TgradT}: the derivative of $\alpha$ can be replaced with an integral by part on the Thomson heat as $ T d \alpha  = d \left( \alpha T \right) - \alpha dT $,
hence the choice of $\alpha(x)$, $\rho(x)$, $\kappa(x)$ for such general cases is straightforward. 
Moreover, as each of $p$- and $n$-legs performance is simulated, the formula can be extended to compute the thermoelectric efficiency of the module with $p$- and $n$-leg pairs: $\eta^{\rm module} = \frac{P(p)+P(n)}{Q_h(p) + Q_h(n)}$, where $p$ and $n$ in parenthesis indicates $p$- and $n$-type legs, respectively.

The \emph{three} average parameters ($\alphaBar$, $\rhoBar$,  $\kappaBar$) and the \emph{three} degrees of freedom ($Z_\gen$, $\tau$, $\beta$) 
are slowlying varying on $I$ and $\gamma$.
For the SnSe \cite{zhao2014ultralow} single-element leg module working at $T_h = 970.1 K$ and $T_c = 302.7 K$, the relative absolute variations for the \emph{six} parameters is less than 0.1\% near the maximum efficiency condition; see Fig. \ref{parameter-variation}.
Even for the segmented leg composed of 0.4 mm low-temperature side segment of BiSbTe \cite{kim2015dense} and 0.6 mm high-temperature side segment of SnSe \cite{zhao2014ultralow} and operating between 970 K and 300 K, calculations reveal that the six parameters are still robust against the change of $I$:
the variation is about 5\% for $\gamma$ in the range of maximum power case and maximum efficiency case.

\begin{figure}
\centering \includegraphics[width=0.8\textwidth]{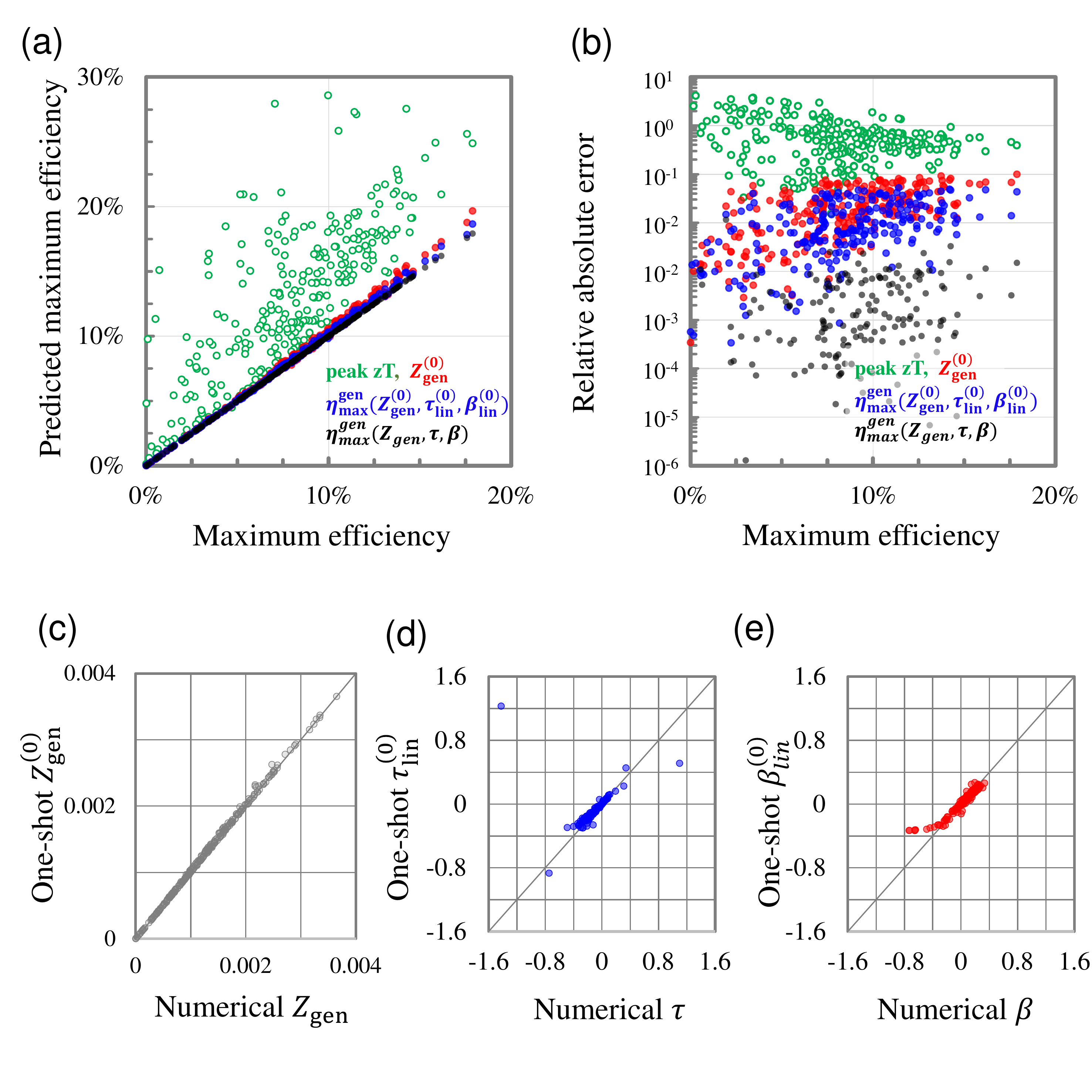}
\caption{
Efficiency estimation methods using $\eta_{\rm max}^{\rm gen} (zT)$,
$\eta_{\rm max}^{\rm gen} (Z_\gen^{(0)})$, 
$\eta_{\rm max}^{\rm gen} (Z_\gen^{(0)}, \tau_{\rm lin}^{(0)}, \beta_{\rm lin}^{(0)})$, and
$\eta_{\rm max}^{\rm gen} (Z_\gen, \tau, \beta | I_{\rm opt})$ are tested for 276 published materials under available temperature range.
(a) Comparison of the efficiency estimation methods and numerically computed maximum efficiency.
(b) Relative absolute errors between the efficiency estimation methods and the numerical maximum efficiency.
(c),(d),(e) Comparison of the one-shot approximation values $Z_\gen^{(0)}$, $\tau_{\rm lin}^{(0)}$, $\beta_{\rm lin}^{(0)}$ in \eqref{one-shot-approx} and the numerical $Z_\gen$, $\tau$, $\beta$ when the maximum efficiency is attained.
}\label{fig-estimation}
\end{figure}

The maximum efficiency can be predicted using the three degrees of freedom ($Z_\gen$, $\tau$, $\beta$).
As the thermoelectric parameters are robust against $I$ and $\gamma$, the $Z_\gen$, $\tau$ and $\beta$ can be assumed to be constants.
To find an approximate value of the maximum efficiency, we maximize the $\eta$ for $\gamma$ fixing $Z_\gen$, $\tau$, $\beta$.
Then we have a general maximum efficiency formula:
\begin{equation}\label{etaMaxGen}
\eta_{\rm max}^{\rm gen} := \frac{\Delta T}{T_h'} \frac{\sqrt{1+Z_\gen T_m'}-1}{\sqrt{1+Z_\gen T_m'}+  T_c' / T_h' },
\end{equation}
where  $T_h' := T_h - \tau \Delta T$, $T_c' := T_c - \left( \tau + \beta \right) \Delta T$, and $T_m' := (T_h' + T_c')/2$.
The maximum efficiency occurs when $\gamma$ is near the $\gamma_{\rm opt}^{\rm gen} :=  \sqrt{1 + Z_\gen T_m' }$.
The computed efficiency results for 276 published thermoelectric material properties with available temperature ranges [see \S1-4 in Supporting Material (SM)] in Fig. \ref{fig-estimation}(a) and (b) validate our general maximum efficiency formula \eqref{etaMaxGen}. It shows the error in efficiency prediction by the single parameter $zT$ is due to the neglect of the hidden parameters $\tau$ and $\beta$.
The general efficiency formula $\eta_{\rm max}^{\rm gen}$ in \eqref{etaMaxGen} is a natural generalization of the traditional effiicency formula of constant property model $\eta_{\rm max}^{\rm const} := \frac{\Delta T}{T_h } \frac{\sqrt{1+zT_m} - 1}{\sqrt{1+zT_m} + T_c / T_h}$ where $z:=\frac{\alpha^2}{\rho\kappa}$ and  $T_m=(T_h+T_c)/2$. Just replacing the $z,T_m,T_h,T_c$ in $\eta_{\rm max}^{\rm const}$ by $Z_\gen,T_m',T_h',T_c'$ respectively, our $\eta_{\rm max}^{\rm gen}$ is obtained. When material properties are constant ($\tau = \beta = 0$), both formulas are the same.

The compatibility factor \cite{snyder2003thermoelectric} can be also generalized.
The optimal current for maximum efficiency $I_{\rm opt} $ is close to $I_{\rm opt}^\gen := \frac{V}{R} \frac{1}{1+\gamma_{\rm max}^{\rm gen}}$.
The compatibility factor $s = \frac{ \sqrt{1+zT}-1}{\alpha T}$ describes the optimal relative current for maximum efficiency at a given $T$.
Hence, a general compatibility factor can be derived as $s_\gen  := \frac{I_{\rm opt}}{K \Delta T} =  \frac{  \sqrt{1 + Z_\gen T_m' }  -1 }{\alphaBar T_m'}$. 

A maximum efficiency formula using the dimensionless weight factors of Joule and Thomson heat $W_J$ and $W_T$ is suggested by Kim et al. \cite{kim2015relationship} However, their formula is not a generalization of the traditional efficiency formula; the $W_T = \frac{\int_{T_c}^{T_h}\int_T^{T_h}T\alpha'(T)\,dT\,dT}{\Delta T\, \int_{T_c}^{T_h}T\alpha'(T)\,dT}$ is not defined when the $\alpha$ is constant because the numerator and denominator vanishes. Also the $W_T$ is not a figure of merit.

While the exact computation of $Z_{\gen}$, $\tau$, $\beta$ require temperature distribution inside the module, an accurate \emph{one-shot approximation} of them is also available (see \S5 in SM):
\begin{equation}\label{one-shot-approx}
\begin{split}
Z_\gen \approx Z_\gen^{(0)} &:= \frac{ \left( \int_{T_c}^{T_h} \alpha(T) \,dT \right)^2 }{\Delta T \,\int_{T_c}^{T_h} \rho (T) \kappa (T) \,dT}, \\
\tau \approx \tau^{(0)}_{\rm lin} &:= -\frac{1}{3} \frac{\alpha_h - \alpha_c}{\alpha_h + \alpha_c}, \\
\beta \approx \beta^{(0)}_{\rm lin} &:= \frac{1}{3}\frac{ \left( \rho\kappa \right)_h -\left( \rho\kappa \right)_c}{ \left( \rho\kappa \right)_h +\left( \rho\kappa \right)_c }.
\end{split}
\end{equation}
Here the subscripts $h$ and $c$ denote the evaluation of the function at $T_h$ and $T_c$ respectively.
The formulas for $Z_\gen^{(0)}$,$\tau^{(0)}_{\rm lin}$,$\beta^{(0)}_{\rm lin}$ in \eqref{one-shot-approx} can be derived using two assumptions: (i) $T=T^{(0)}$ where $T^{(0)}$ is the temperature distribution for the $J=0$ case, i.e. $T^{(0)}$ is a solution of $-\kappa(T^{(0)}) \frac{dT^{(0)}}{dx} = \text{const}$. The superscript ${(0)}$ means we use the $J=0$ case. (ii) the material properties $\alpha$ and $\rho \times \kappa$ are linear with respect to $T$. The subscripts in $\tau^{(0)}_{\rm lin}$ and $\beta^{(0)}_{\rm lin}$ emphasize the linearity.
In Fig. \ref{fig-estimation}(c)--(e), the strong correlation between the one-shot approximation values and exact numerical values is verified for the 276 thermoelectric material properties from literature. The approximation fails only when the Seebeck coefficient is small and sign-changing with temperature.

The one-shot approximation \eqref{one-shot-approx} clarifies the effect of temperature-dependent material properties on the efficiency. For example, because $\tau \approx -\frac{1}{3} \frac{\alpha_h - \alpha_c}{\alpha_h + \alpha_c}$, the efficiency can be enhanced if the $\alpha(T)$ declines more rapidly on $T$, i.e. the Thomson effect as a heat sink becomes stronger.
Hence the Thomson effect in efficiency estimation is important as reported previously 
\cite{sherman1960calculation, sunderland1964influence, ybarrondo1965influence, min2004thermoelectric, wee2011analysis, kim2015relationship}.

\begin{figure}
\centering\includegraphics[width=0.8\textwidth]{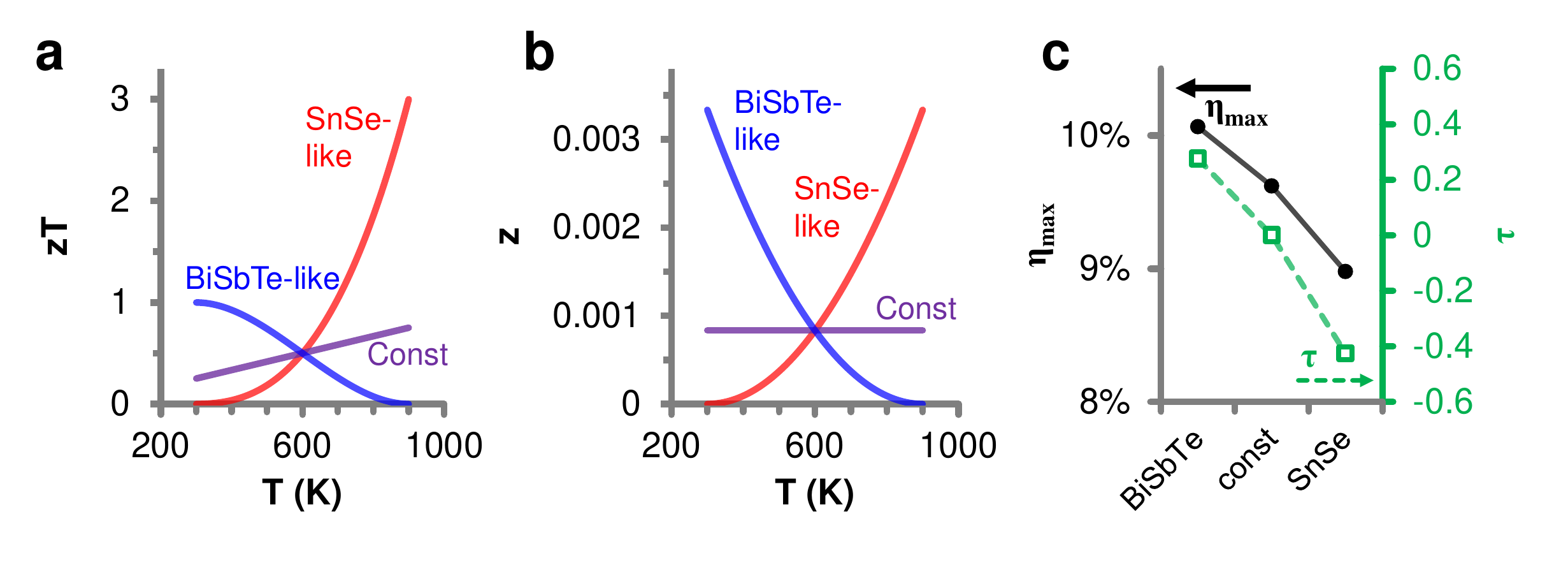}
\caption{(a) The $zT$, (b) the $z$, (c) the maximum efficiency and $\tau$ for three imaginary materials which imitates {\rm BiSbTe}-like, {\rm SnSe}-like, and constant-$z$ materials. The $\alpha$ of the materials is linear while the $\rho$ and $\kappa$ of them are constant. The materials have the same $\alphaBar$ and $Z_\gen$. For working temperature range from $300 K$ to $900 K$, the highest maximum efficiency is found in the {\rm BiSbTe}-like material due to the positive $\tau$.
}\label{fig-eg-zT-fail}
\end{figure}

The one-shot approximation \eqref{one-shot-approx} also explains why the single parameter generalizations of the $zT$ fail for efficiency prediction of some materials.
For example, the peak $zT$ of {\rm SnSe} materials is significantly greater than that of {\rm BiSbTe} materials ({\rm SnSe} has the highest peak $zT$ of 2.6 at 923 K \cite{zhao2014ultralow}), but the efficiency of {\rm BiSbTe} is significantly greater than that of the {\rm SnSe}. This extreme failure of $zT$ can be explained by the figure of merit $\tau$. Consider three imaginary materials imitating {\rm BiSbTe}-like, {\rm SnSe}-like, and constant-$z$ materials. For simplicity, we impose two assumptions on their material properties: (i) the $\rho$ and $\kappa$ of them are $T$-independent and they have the same $\alphaBar$. (ii) the $\alpha$ of them is linear on temperature; the {\rm BiSbTe}-like material has linearly decreasing $\alpha$, the {\rm SnSe}-like material has linearly increasing $\alpha$, and the constant-$z$ material has the constant $\alpha$. Then, as shown in Fig. \ref{fig-eg-zT-fail}, the peak $zT$ of the {\rm SnSe}-like material is very high as the $zT$ is proportional to $T^3$. However, due to the temperature-dependent profile of $\alpha$, the $\tau$ of the {\rm SnSe}-like material is \emph{negative} while the $\tau$ of {\rm BiSbTe}-like material is positive; see \eqref{tau-beta} and \eqref{one-shot-approx}. Since the $Z_\gen$ is the same for the three materials, the $\tau$ is the main figure of merit which concludes that the $\rm BiSbTe$-like material has higher maximum efficiency than the $\rm SnSe$-like material. This example shows a single average parameter is not enough for efficiency prediction and the gradient of material properties can be important.

In summary, 
three degrees of freedom $Z_\gen$, $\tau$ and $\beta$ exactly determine the thermoelectric conversion efficiency. Each degree of freedom is a figure of merit hence improving $\tau$ or $\beta$ is a complementary way to increase the efficiency.
The physical insights on high $\tau$ or $\beta$ materials explain why single parameter approaches are not enough for some materials, and can be used to evaluate and optimize the thermoelectric materials and devices.

\section*{Acknowledgement}
B.R. and J.C. contributed equally to this work. This work was supported by KERI primary research program through the NST funded by the MSIT of the Republic of Korea (ROK): grant No. 19-12-N0101-22. It was also supported by the KETEP and the MOTIE of the ROK: 
grant Nos. 20162000000910, 20172010000830, 20188550000290.

\title{\textcolor{blue}{\bf{Supplementary Material for \\    ``General Efficiency Theory of Thermoelectric Conversion''}}}

\begin{abstract}
This Suppelmentary Material (SM) is prepared to support the paper for publication in the \emph{Applied Physics Letters}, entitled \emph{General Efficiency Theory of Thermoelectric Conversion}. This SM is composed of following sections.
In \textcolor{blue}{\bf \S1}, we give information on \emph{full reference list for 276 thermoelectric materials} from 263 literatures, which were used for efficiency prediction (Fig. 3)
In \textcolor{blue}{\bf \S2}, we describe \emph{how to compute} numerical ideal maximum efficiency for thermoelectric conversion of single-leg materials in Fig. 3. 
In \textcolor{blue}{\bf \S3}, we give a statistical analysis for \emph{the predicted efficiency of 276 materials} (Fig. 3).
In \textcolor{blue}{\bf \S4}, we give an efficiency analysis data for \emph{selective 18 thermoelectric materials}.
In \textcolor{blue}{\bf \S5}, we give a full derivation of one-shot approximation forms for $Z_\gen$, $\tau$, and $\beta$.
\end{abstract}

\section*{\bf{\S1 Thermoelectric Property (TEP) Data used in Figure 2}}
In this work, we constructed a dataset of TEPs of 276 materials gathered from 263 literatures \cite{biswas_strained_2011, biswas_high-performance_2012, fu_realizing_2015, gelbstein_controlling_2013, he_ultrahigh_2015, heremans_enhancement_2008, hsu_cubic_2004, hu_shifting_2014, hu_power_2016, kim_dense_2015, lin_tellurium_2016, liu_thermoelectric_2011, liu_convergence_2012, pan_thermoelectric_2016, pei_convergence_2011, pei_high_2011-1, poudel_high-thermoelectric_2008, rhyee_peierls_2009, wang_right_2014, zhao_raising_2012, zhao_thermoelectrics_2012, zhao_ultralow_2014, zhao_ultrahigh_2015, cui_thermoelectric_2007, cui_crystal_2008, eum_transport_2015, fan_p-type_2010, han_alternative_2013, hsu_enhancing_2014, zheng_mechanically_2014, hu_tuning_2015, hwang_enhancing_2013, ko_nanograined_2013, zhang_improved_2015, zhao_bismuth_2005, lee_control_2010, lee_enhancement_2013, lee_crystal_2013, yan_experimental_2010, lee_preparation_2013, lee_preparation_2014, lee_preparation_2014-1, lee_preparation_2014-2, lee_thermoelectric_2014, sumithra_enhancement_2011, lukas_transport_2012, min_surfactant-free_2013, mun_fe-doping_2015, ovsyannikov_enhanced_2015, puneet_preferential_2013, shin_twin-driven_2014, son_n-type_2012, son_effect_2013, soni_enhanced_2012, xiao_enhanced_2014, tang_preparation_2007, wang_metal_2013, wu_thermoelectric_2013, yelgel_thermoelectric_2012, zhang_rational_2012, wei_minimum_2016, lan_high_2012, yu_preparation_2013, kosuga_enhanced_2014, scheele_thermoelectric_2011, ahn_improvement_2009, ahn_exploring_2010, ahn_enhanced_2013,androulakis_thermoelectric_2010, androulakis_high-temperature_2011, bali_thermoelectric_2013, bali_thermoelectric_2014, wu_superior_2015, dong_transport_2009, dow_effect_2010, falkenbach_thermoelectric_2013, fan_enhanced_2015, fang_synthesis_2013, jaworski_valence-band_2013, jian_significant_2015, keiber_complex_2013, kim_spinodally_2016, lee_improvement_2012, lee_contrasting_2014, li_enhanced_2013, li_pbte-based_2014, liu_effect_2013, lo_phonon_2012, lu_enhancement_2013, pei_combination_2011, pei_high_2011, pei_self-tuning_2011, pei_stabilizing_2011, pei_low_2012, pei_thermopower_2012, pei_optimum_2014, poudeu_high_2006, rawat_thermoelectric_2013, wang_large_2013, wang_tuning_2014, wu_broad_2014, wu_strong_2014, yamini_heterogeneous_2015, yang_enhanced_2015, zebarjadi_power_2011, zhang_enhancement_2012, zhang_heavy_2012, zhang_effect_2013, zhang_enhancement_2015, al_rahal_al_orabi_band_2015, banik_mg_2015, banik_high_2016, banik_agi_nodate, chen_thermoelectric_2014, chen_understanding_2016, leng_thermoelectric_2016, pei_interstitial_2016, tan_high_2014, tan_codoping_2015, tan_valence_2015, tang_realizing_2016, wang_thermoelectric_2015, zhang_high_2013, zhou_optimization_2014, guan_thermoelectric_2015, suzuki_supercell_2015, fahrnbauer_high_2015, gelbstein_-doped_2007, gelbstein_powder_2007, gelbstein_thermoelectric_2010, hazan_effective_2015, kusz_structure_2016, lee_influence_2014, schroder_nanostructures_2014, schroder_tags-related_2014, williams_enhanced_2015, wu_origin_2014, aikebaier_effect_2010, chen_thermoelectric_2012, dow_thermoelectric_2009, drymiotis_enhanced_2013, du_effect_2014, guin_sb_2015, han_lead-free_2012, he_synthesis_2012, hong_anomalous_2014, liu_enhanced_2016, mohanraman_influence_2014, pei_alloying_2011, wang_synthesis_2008, wu_state_2015, zhang_improved_2010, aizawa_solid_2006, akasaka_composition_2007, akasaka_non-wetting_2007, cheng_mg2si-based_2016, duan_effects_2016, kajikawa_thermoelectric_1998, liu_n-type_2015, luo_fabrication_2009, mars_thermoelectric_2009, noda_temperature_1992, tani_thermoelectric_2005, tani_thermoelectric_2007-1, tani_thermoelectric_2007, yang_preparation_2009, yin_optimization_2016, zhang_high_2008, zhang_situ_2008, zhang_suppressing_2015, zhao_synthesis_2009, joshi_enhanced_2008, tang_holey_2010, wang_enhanced_2008, ahn_improvement_2012, bhatt_thermoelectric_2014, fu_band_2015, kraemer_high_2015, krez_long-term_2015, liu_thermoelectric_2007, mudryk_thermoelectricity_2002, shi_low_2008, bai_enhanced_2009, bao_effect_2009, chitroub_thermoelectric_2009, dong_hpht_2009, duan_synthesis_2012, dyck_thermoelectric_2002, he_thermoelectric_2007, he_great_2008, laufek_synthesis_2009, li_thermoelectric_2005, liang_ultra-fast_2014, liu_enhanced_2007, mallik_transport_2008, mallik_thermoelectric_2013, mi_thermoelectric_2008, pei_thermoelectric_2008, qiu_high-temperature_2011, rogl_thermoelectric_2010, rogl_new_2011, rogl_n-type_2014, rogl_new_2015, sales_filled_1996, shi_multiple-filled_2011, stiewe_nanostructured_2005, su_structure_2011, tang_synthesis_2001, xu_thermoelectric_2014, yang_synthesis_2009, zhang_high-pressure_2012, zhao_enhanced_2009, zhou_thermoelectric_2013, bali_thermoelectric_2016, ding_high_2016, jo_simultaneous_2016, joo_thermoelectric_2016, li_enhanced_2016, li_inhibition_2016, liu_enhanced_2012, zhou_strategy_2017, zhou_scalable_2017, zhang_discovery_2017, xu_nanocomposites_2017, xie_stabilization_2016, wang_high_2016, seo_effect_2017, pei_multiple_2016, park_extraordinary_2016, moon_tunable_2016, zhu_nanostructuring_2007, choi_thermoelectric_1997, yamanaka_thermoelectric_2003, yang_nanostructures_2008, yang_natural_2010, zhang_effects_2009, zhou_nanostructured_2008, sharp_properties_2003, salvador_transport_2009, levin_analysis_2011, yamanaka_thermoelectric_2003-1, zhao_thermoelectric_2008, zhao_synthesis_2006, yu_high-performance_2009, xiong_high_2010, toberer_traversing_2008, chung_csbi4te6:_2000, tang_high_2008, mi_improved_2007, liu_improvement_2008, li_preparation_2008, chen_high_2006, zhong_high_2014, yu_thermoelectric_2012, liu_ultrahigh_2013, liu_copper_2012, he_high_2015, gahtori_giant_2015, day_high-temperature_2014, ballikaya_thermoelectric_2013, bailey_enhanced_2016, li_promoting_2017, liang_enhanced_2017} to test our method.
To digitize the TEP data, we use the Plot Digitizer \cite{PlotDigitizer}.
The dataset consists of Seebeck coefficient $\alpha$, electrical resistivity $\rho$ (or electrical conductivity $\sigma$), and thermal conductivity $\kappa$ at measured temperature $T$. For the numerical computation of efficiency, we use the available temperature ranges of the given material: the $T_c$ is defined as the maximum of the lowest measured temperautre and $T_h$ is defined as the minimum of the highest measured temperature for given materials.

As shown in Table \ref{tep-dataset}, 
the 276 materials in our dataset have various base-material groups: 59 $\rm Bi_2Te_3$-related materials, 55 $\rm PbTe$-related materials , 40 skutterudite (SKD), 23 $\rm Mg_2Si$-based materials, 18 $\rm GeTe$ materials, 14 $\rm M_2Q$ antifluorite-type chalcogenide materials (where M = Cu, Ag, Au and Q = Te, Se), 12 $\rm SnTe$-related materials, 11 $\rm ABQ_2$-type materials (where A=Group I, B=Bi, Sb, Q=Te, Se), 8 $\rm SnSe$-related materials, 7 $\rm PbSe$-related materials, 7 half-Heusler (HH) materials, 6 $\rm SiGe$-related materials, 3 $\rm In_4 Se_3$-related materials, 3 $\rm PbS$-related materials, 2 oxide materials, 2 clathrate materials, and 6 others. Here the base-material denotes the representative material, not the exact composition. Also note that for the categorizatoin of base materials, the doping element is ignored. For examples, $\rm Bi_2Te_3$, $\rm Sb_2Te_3$, $\rm Bi_2Se_3$ binary and their ternary alloys are categorized as $\rm Bi_2 Te_3$-related materials. The material doping composition is not denoted in the composition of the base material.

\begin{table}[h]
\caption{TEP Dataset of 276 materials with various material groups. `Group' and `\#mat.' coloumns represent the group of base material and the number of materials inside the Group.}
\label{tep-dataset}
\begin{tabular}{|c|c|c|c|}
\hline
\textbf{Group} & \textbf{\#mat.} & \textbf{Group} & \textbf{\#mat.} \\ \hline
$\rm Bi_2Te_3$ & 59 & SnSe & 8 \\ \hline
PbTe & 55 & PbSe & 7 \\ \hline
SKD & 40 & HH & 7 \\ \hline
$\rm Mg_2Si$ & 23 & SiGe & 6 \\ \hline
GeTe & 18 & $\rm In_4Se_3$ & 3 \\ \hline
$\rm M_2Q$ & 14 & PbS & 3 \\ \hline
SnTe & 12 & Oxide & 2 \\ \hline
$\rm ABQ_2$ & 11 & clathrate & 2 \\ \hline
etc. & 6 & \textbf{Total} & \textbf{276} \\ \hline
\end{tabular}
\end{table}

\section*{\bf{\S2 Numerical Efficiency Calculation in Figure 3}}
Thermoelectric phenomena in power module is governed by the thermoelectric differential equation as below:
\begin{equation}\label{SM_TEQ-PDE-1D}
\frac{d}{dx} \left(\kappa \frac{dT}{dx} \right) -T \frac{d \alpha}{dT} \frac{dT}{dx} J +  \rho J^2  = 0.
\end{equation}

Numerical maximum efficiencies of ideal thermoelectric modules without thermal loss by radiation or air convection are computed for 276 materials and compared with the peak $zT$ values.
The thermoelectric properties are fitted using the \emph{piecewise linear interpolation} at intermediate temperatures.
The exact temperature distribution $T(x)$ of steady state is determined by solving the differential equations of thermoelectricity (see equation-\eqref{SM_TEQ-PDE-1D}) with Dirichlet boundary conditions; the end point temperature is determined from the available temperature range.
Then the thermoelectric performances of a thermoelectric leg with length $L$ and cross sectional area $A$ are calculated as a function of current density $J$ given as
$\eta (J) = \frac{P/A}{Q_h/A} = \frac{ J (   \int_c^h \alpha dT -  J \int_0^L \rho dx ) }{ -   \kappa_h \nabla T_h + J \alpha_h T_h } $, where the $P$ and $Q_h$ are the power delivered outside and the hot-side heat current respectively. Then, the maximum of numerical efficiency ($\eta_{\rm max}$) is calculated, which satisfies the relation $\eta (J)  \leq \eta_{\rm max}$.
The reduced efficiency $\eta_{\rm red}$ is obtained as $\eta_{\rm red} = \frac{\eta_{\rm max}}{\eta_{\rm Carnot}}$, where $\eta_{\rm Carnot} = \frac{T_h - T_c}{T_h}$.

\clearpage
\section*{\bf{\S3 Maximum efficiency prediction using $\eta_{\rm max}^{\rm gen}$ in Figure 3 }}
In Figure 3(b), we can observe that the maximum efficiency estimation formula $\eta_{\rm max}^{\rm gen} (Z_\gen, \tau, \beta)$ in equation (8) is highly accurate.
In Table \ref{Table-rel-err-eff-formula}, various statistics on the relative error of maximum efficiency ($\frac{\eta_{\rm max}^{\rm gen}  -  \eta_{\rm max}}{ \eta_{\rm max}  }$) are given.
\begin{table}[h]
\caption{Statistics on the relative error (RelErr) of the maximum efficiency estimation formula $\eta_{\rm max}^{\rm gen} (Z_\gen, \tau, \beta)$. Average (Avg), root mean square (RMS RelErr or StdErr), maximum (max), and minimum (min) of the relative errors are estimated for 276 materials for thermoelectric power generator working at their available temperature.} \label{Table-rel-err-eff-formula}
\begin{tabular}{|c|r|r|r|r|r|r|}
\hline
\multirow{3}{*}{\begin{tabular}[c]{@{}c@{}}276 materials\\ for power\\ module\end{tabular}} & \multicolumn{6}{c|}{Relative error in maximum efficiency formula} \\ \cline{2-7} 
 & \multicolumn{5}{c|}{$\eta_{\rm max}^{\rm gen}$} & \multicolumn{1}{c|}{$\eta_{\rm max}^{\rm const}$} \\ \cline{2-7} 
 & \multicolumn{1}{c|}{$Z_\gen,\tau,\beta$} & \multicolumn{1}{c|}{$Z_\gen^{(0)},\tau^{(0)},\beta^{(0)}$} & \multicolumn{1}{c|}{$Z_\gen^{(0)},\tau_{\rm lin}^{(0)},\beta_{\rm lin}^{(0)}$} & \multicolumn{1}{c|}{$Z_\gen$} & \multicolumn{1}{c|}{$Z_\gen^{(0)}$} & \multicolumn{1}{c|}{peak $zT$} \\ \hline
Avg RelErr & 0.02\% & 1.11\% & 1.08\% & 1.42\% & 2.29\% & 235\% \\ \hline
StdErr (RMS RelErr) & 0.09\% & 1.38\% & 1.38\% & 1.52\% & 2.47\% & 1854\% \\ \hline
max RelErr & 1.15\% & 5.45\% & 5.23\% & 5.80\% & 9.96\% & 28835\% \\ \hline
min RelErr & -0.61\% & -1.92\% & -1.76\% & -1.78\% & -2.48\% & -4\% \\ \hline
\end{tabular}
\end{table}

If we use the exact $Z_\gen, \tau, \beta$, the standard error (=root mean square of relative errors) of $\eta_{\rm max}^{\rm gen}$ is $9.60 \times 10^{-4}$. Actually, these small value is generated during numerical analysis.

If we use $Z_\gen^{(0)}, \tau_{\rm lin}^{(0)}, \beta_{\rm lin}^{(0)}$, the standard error is $1.75 \times 10^{-2}$. 
For the signle crystalline $\rm SnSe$ with peak $zT$ of 2.6, the relative error of one shot method is found to be only $6.82 \times 10^{-3}$.
However, when we use the different approximation such as linear $T(x)$ or different average scheme for $z$, the error becomes larger than ours due to the non-linearity of $T$ for this material \cite{kim2015relationship}.

If we only use the $Z_\gen^{(0)}$ with zero $\tau$ and $\beta$, 
the efficiency is still well predicted with the standard error of  $3.37 \times 10^{-2}$.
But, in some materials, the error is relatively large due to the neglect of the $\tau$ and $\beta$. The largest relative error of 10\% is found for \cite{wu_broad_2014}, due to the non-vanishing gradient parameters 
($\tau = -0.222 \approx \tau^{(0)} = -0.177 \approx  \tau_{\rm lin}^{(0)} = -0.204$,
$\beta = 0.2085 \approx \beta^{(0)} = 0.228 \approx  \beta_{\rm lin}^{(0)} = 0.185$, when $T_h = 918 K$ and $T_c = 304 K$).

\section*{\bf{\S4 Efficiency analysis for selective 18 TE materials}}

\begin{figure}
\centering \includegraphics[width=\textwidth]{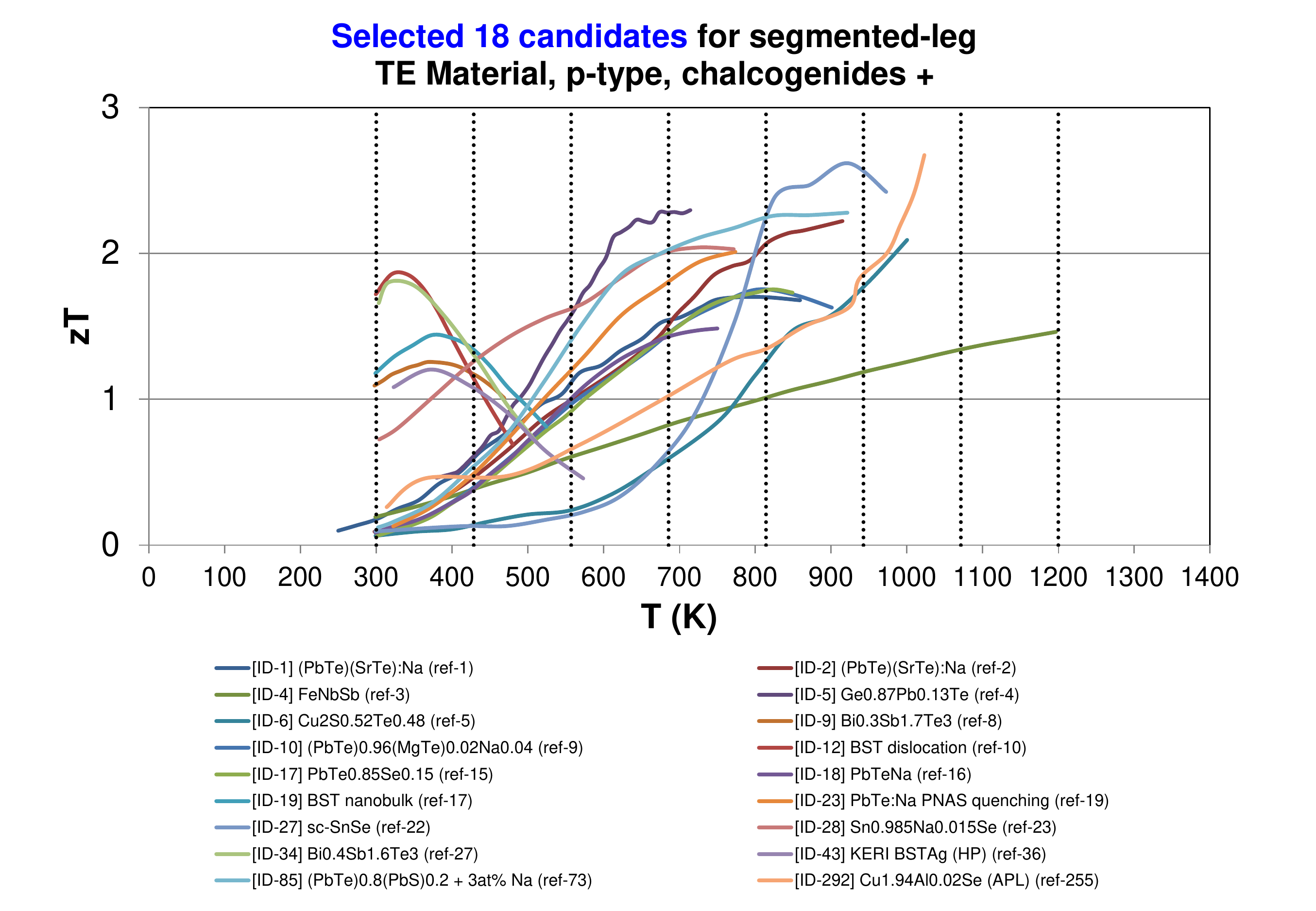}
\caption{The $zT$ curves for 18 selected materials. The `ref-\#' is the reference number.}\label{zT-for-18-mats}
\end{figure}

As the representative, we consider 18 thermoelectric materials showing high peak $zT$ values exceeding 1. The TEP curves for temperature dependent Seebeck coefficient, electrical conductivity, and thermal conductivity can be found in the additional excel SM file. The full $zT$ curves of them are shown in Figure \ref{zT-for-18-mats}.
Table \ref{table-18-mats-temp-range}, \ref{table-18-mats-max-eff} and \ref{table-18-mats-te-dof} contain more information of the materials, including available temperature range, peak $zT$, numerical efficiency, formula efficiency, and the thermoelectric degrees of freedom.

\begin{table}[h]
\caption{Information of 18 selected materials: available temperature range $T_c$ and $T_h$, $\Delta T = T_h - T_c$, peak $zT$, temperature of the peak $zT$.}\label{table-18-mats-temp-range}
\begin{tabular}{|l|l|l|l|l|}
\hline
ID-\# & Material or Process [Reference] & $T_c$ (K) & $T_h$ (K) & peat-$zT$ @$T$ \\ \hline
ID-1 & (PbTe)(SrTe):Na  \cite{biswas_strained_2011} & 251 & 818 & 1.7  @800K \\ \hline
ID-2 & (PbTe)(SrTe):Na  \cite{biswas_high-performance_2012} & 302 & 915 & 2.2  @915K \\ \hline
ID-4 & FeNbSb  \cite{fu_realizing_2015} & 301 & 1200 & 1.5  @1200K \\ \hline
ID-5 & $\rm Ge_{0.87}Pb_{0.13}Te$  \cite{gelbstein_controlling_2013} & 329 & 713 & 2  @673K \\ \hline
ID-6 & $\rm Cu_2S_{0.52}Te_{0.48}$  \cite{he_ultrahigh_2015} & 299 & 997 & 2.1  @1000K \\ \hline
ID-9 & $\rm Bi_{0.3}Sb_{1.7}Te_3$  \cite{hu_shifting_2014} & 298 & 479 & 1.3  @380K \\ \hline
ID-10 & $\rm (PbTe)_{0.96}(MgTe)_{0.02}Na_{0.04}$  \cite{hu_power_2016} & 307 & 900 & 1.8  @810K \\ \hline
ID-12 & BST  dislocation \cite{kim_dense_2015} & 300 & 480 & 1.86  @320K \\ \hline
ID-17 & $\rm PbTe_{0.85}Se_{0.15}$  \cite{pei_convergence_2011} & 300 & 847 & 1.8  @850K \\ \hline
ID-18 & PbTeNa  \cite{pei_high_2011-1} & 300 & 750 & 1.4  @750K \\ \hline
ID-19 & BST nanobulk  \cite{poudel_high-thermoelectric_2008} & 300 & 525 & 1.4  @373K \\ \hline
ID-23 & PbTe:Na, quenching (PNAS) \cite{wang_right_2014} & 321 & 759 & 2  @773K \\ \hline
ID-27 & sc-SnSe, $b$-axis  \cite{zhao_ultralow_2014} & 303 & 970 & 2.6  @923K \\ \hline
ID-28 & $\rm Sn_{0.985}Na_{0.015}Se$  \cite{zhao_ultrahigh_2015} & 304 & 773 & 2  @773K \\ \hline
ID-34 & $\rm Bi_{0.4}Sb_{1.6}Te_3$  \cite{fan_p-type_2010} & 303 & 513 & 1.8  @316K \\ \hline
ID-43 & KERI BSTAg, HP  \cite{lee_control_2010} & 323 & 573 & 1.2  @373K \\ \hline
ID-85 & $\rm (PbTe)_{0.8}(PbS)_{0.2}$ + 3at\% Na \cite{wu_superior_2015} & 302 & 922 & 2.3  @923K \\ \hline
ID-292 & $\rm Cu_{1.94}Al_{0.02}Se$ (APL)  \cite{zhong_high_2014} & 327 & 1019 & 2.62  @1029K \\ \hline
\end{tabular}
\end{table}

\begin{table}[h]
\caption{Information of 18 selected materials: (a) maximum efficiencies computed using exact numerical method ($T$ is computed by fixed-point interation, then power, heat and efficiency are computed), 
maximum efficiencies computed from general maximum efficiency formula $\eta_{\rm max}^{\rm gen}$ (see equation (8) in the Manuscript) 
(b) using \emph{exact} thermoeletric degrees of freedom (DOFs) with exact $T$ ($Z_\gen,\tau,\beta$), 
(c) using DOFs with $T^{(0)}$ ($Z_\gen^{(0)},\tau^{(0)},\beta^{(0)}$), 
(d) using DOFs with \emph{one-shot} approximation ($Z_\gen^{(0)},\tau_{\rm lin}^{(0)},\beta_{\rm lin}^{(0)}$),
(e) using DOFs with only $Z_\gen$ while $\tau=\beta=0$,
(f) using DOFs with only $Z_\gen^{(0)}$ while $\tau=\beta=0$,
and (g) using the classical efficiency formula for constant TEP using peak $zT$.
Note that when we computing the numerical maximum efficiency we calculate the $T$ using the fixed-point iteration with integral equation of $T$ for given $J$. Then $J$ is optimized to maximize the efficiency. To compute $Z_\gen$, $\tau$, and $\beta$, we used the $T$ distribution of the $J$ of the maximum efficiency condition. For one shot approximations, we use the equation (9) in the Manuscript.
}
\label{table-18-mats-max-eff}
\begin{tabular}{|l|c|c|c|c|c|c|c|}
\hline
\multirow{3}{*}{ID-\#} & \multicolumn{7}{c|}{$\eta_{\rm max}$} \\ \cline{2-8} 
 & \multirow{2}{*}{\makecell{(a)\\ exact}} & \multicolumn{5}{c|}{$\eta_{\rm max}^{\rm gen}$} 
 &  $\eta_{\rm max}^{\rm const}$ \\ \cline{3-8} 
 &  
 & \makecell{(b)\\ $Z_\gen,\tau,\beta$}
 & \makecell{(c)\\ $Z_\gen^{(0)},\tau^{(0)},\beta^{(0)}$} 
 & \makecell{(d)\\ $Z_\gen^{(0)},\tau_{\rm lin}^{(0)},\beta_{\rm lin}^{(0)}$}
 & \makecell{(e)\\ $Z_\gen$}
 & \makecell{(f)\\ $Z_\gen^{(0)}$}
 & \makecell{(g)\\ peak $zT$} \\ \hline
ID-1 & 13.7\% & 13.7\% & 14.4\% & 14.3\% & 14.5\% & 15\% & 22.9\% \\ \hline
ID-2 & 15.9\% & 15.9\% & 16.2\% & 16.1\% & 16.6\% & 16.8\% & 24.9\% \\ \hline
ID-4 & 15.3\% & 15.3\% & 15.8\% & 15.8\% & 15.8\% & 16.3\% & 23.8\% \\ \hline
ID-5 & 12.5\% & 12.6\% & 12.9\% & 13\% & 13.1\% & 13.4\% & 18\% \\ \hline
ID-6 & 10.5\% & 10.5\% & 10.7\% & 10.7\% & 11.1\% & 11.1\% & 25.9\% \\ \hline
ID-9 & 8.4\% & 8.4\% & 8.4\% & 8.4\% & 8.4\% & 8.4\% & 9.2\% \\ \hline
ID-10 & 13.8\% & 13.8\% & 14.2\% & 14.1\% & 14.4\% & 14.7\% & 22\% \\ \hline
ID-12 & 9.1\% & 9.1\% & 9.1\% & 9.1\% & 9\% & 9\% & 11.2\% \\ \hline
ID-17 & 12.6\% & 12.7\% & 13\% & 12.9\% & 13.3\% & 13.5\% & 21.5\% \\ \hline
ID-18 & 10.4\% & 10.4\% & 10.8\% & 10.8\% & 10.9\% & 11.2\% & 16.9\% \\ \hline
ID-19 & 9.9\% & 9.9\% & 10\% & 10\% & 9.9\% & 9.9\% & 11.1\% \\ \hline
ID-23 & 11.6\% & 11.6\% & 12.1\% & 12.1\% & 12.2\% & 12.5\% & 19.6\% \\ \hline
ID-27 & 7.1\% & 7.1\% & 7.1\% & 7.1\% & 7.1\% & 7.1\% & 27.9\% \\ \hline
ID-28 & 16.2\% & 16.2\% & 16.9\% & 16.9\% & 16.7\% & 17.3\% & 20.9\% \\ \hline
ID-34 & 10.1\% & 10.1\% & 10.1\% & 10.1\% & 10\% & 10\% & 12.2\% \\ \hline
ID-43 & 8.2\% & 8.2\% & 8.2\% & 8.2\% & 8.1\% & 8.1\% & 10.3\% \\ \hline
ID-85 & 17.6\% & 17.6\% & 18.1\% & 17.8\% & 18.5\% & 18.8\% & 25.6\% \\ \hline
ID-292 & 14.3\% & 14.3\% & 14.9\% & 14.9\% & 14.9\% & 15.4\% & 27.5\% \\ \hline
\end{tabular}
\end{table}

\begin{table}[h]
\caption{Information of 18 selected materials: \emph{exact} value and \emph{one-shot} approximation of thermoeletric degrees of freedom.}
\label{table-18-mats-te-dof}
\begin{tabular}{|l|r|r|r|r|r|r|}
\hline
\multicolumn{1}{|c|}{ID-\#} & \multicolumn{1}{c|}{$Z_\gen$} & \multicolumn{1}{c|}{$\tau$} & \multicolumn{1}{c|}{$\beta$} & \multicolumn{1}{c|}{$Z_\gen^{(0)}$} & \multicolumn{1}{c|}{$\tau_{\rm lin}^{(0)}$} & \multicolumn{1}{c|}{$\beta_{\rm lin}^{(0)}$} \\ \hline
ID-1 & 0.0015 & -0.253 & 0.192 & 0.0016 & -0.207 & 0.199 \\ \hline
ID-2 & 0.0018 & -0.186 & 0.068 & 0.0018 & -0.152 & 0.074 \\ \hline
ID-4 & 0.0010 & -0.164 & 0.197 & 0.0011 & -0.141 & 0.203 \\ \hline
ID-5 & 0.0022 & -0.227 & 0.094 & 0.0023 & -0.168 & 0.105 \\ \hline
ID-6 & 0.0008 & -0.253 & 0.027 & 0.0008 & -0.208 & 0.028 \\ \hline
ID-9 & 0.0029 & -0.019 & 0.135 & 0.0029 & -0.017 & 0.136 \\ \hline
ID-10 & 0.0015 & -0.192 & 0.102 & 0.0015 & -0.161 & 0.107 \\ \hline
ID-12 & 0.0033 & 0.030 & 0.177 & 0.0033 & 0.032 & 0.178 \\ \hline
ID-17 & 0.0014 & -0.231 & 0.109 & 0.0015 & -0.189 & 0.112 \\ \hline
ID-18 & 0.0014 & -0.271 & 0.167 & 0.0014 & -0.214 & 0.172 \\ \hline
ID-19 & 0.0028 & -0.015 & 0.189 & 0.0028 & -0.013 & 0.190 \\ \hline
ID-23 & 0.0017 & -0.254 & 0.138 & 0.0017 & -0.194 & 0.142 \\ \hline
ID-27 & 0.0005 & 0.082 & -0.379 & 0.0005 & 0.086 & -0.382 \\ \hline
ID-28 & 0.0025 & -0.154 & 0.217 & 0.0026 & -0.118 & 0.225 \\ \hline
ID-34 & 0.0032 & 0.033 & 0.164 & 0.0032 & 0.036 & 0.166 \\ \hline
ID-43 & 0.0019 & 0.028 & 0.186 & 0.0019 & 0.029 & 0.187 \\ \hline
ID-85 & 0.0021 & -0.179 & 0.079 & 0.0021 & -0.146 & 0.095 \\ \hline
ID-292 & 0.0013 & -0.211 & 0.178 & 0.0014 & -0.166 & 0.187 \\ \hline
\end{tabular}
\end{table}

\clearpage
\section*{\bf{\S5 One-shot approximation $Z_\gen^{(0)}$, $\tau_{\rm lin}^{(0)}$ and $\beta_{\rm lin}^{(0)}$}}\label{sec-one-shot}
The exact forms of $Z_\gen$, $\tau$ and $\beta$ are written as
\begin{equation}\label{Zgen}
Z_\gen := \frac{(V/\Delta T)^2}{RK} = \frac{\alphaBar^2}{\rhoBar \,\kappaBar},
\end{equation}
\begin{align}
\tau &:= \frac{1}{\alphaBar \Delta T} \left[ (\alphaBar - \alpha_h) T_h -K \,\delta T^{(1)} \right] \label{SI-tau},\\
\beta &:= \frac{2}{R} K \,\delta T^{(2)} - 1. \label{SI-beta}
\end{align}

The computation of $Z_\gen$, $\tau$ and $\beta$ requires the exact temperature distribution.
But they can be estimated directly from the material properties. In this section we derive an approximate formula for $Z_\gen$, $\tau$ and $\beta$.
The idea is to use the temperature distribution for $J=0$, which is similar to the exact temperature distribution because most devices induce small $J$ due to the small $zT$. Let $T^{(0)}$ be the temperature distribution for $J=0$ and define

\begin{alignat*}{3}
\rhoBar^{(0)} &:= \frac{1}{L} \int_{0}^{L} \rho(T^{(0)}(x)) \,dx &=& \frac{A}{L} R^{(0)}, \\
\frac{1}{\kappaBar^{(0)}} &:= \frac{1}{L} \int_{0}^{L} \frac{1}{\kappa (T^{(0)}(x))} \,dx &=& \frac{A}{L} \frac{1}{K^{(0)}}.
\end{alignat*}
From the thermoelectric differential equation \eqref{SM_TEQ-PDE-1D} with $J=0$, we can check that
\begin{equation} \label{SI-heat-flux-approx}
-\kappa(T^{(0)}(x)) \frac{d T^{(0)}}{dx}(x) = \kappaBar^{(0)} \frac{\Delta T}{L}.
\end{equation}
Hence
\[
\begin{split}
\int_{T_c}^{T_h} \rho(T) \kappa(T)\,dT &= \int_{T_c}^{T_h} \rho(T^{(0)}) \Big(-\frac{\Delta T}{L} \kappaBar^{(0)}\Big) \frac{dx}{dT^{(0)}} \,dT^{(0)} \\
&= \frac{\Delta T}{L} \int_0^L \rho(T^{(0)}(x)) \,\kappaBar^{(0)} \,dx \\
&= \Delta T \,\rhoBar^{(0)} \,\kappaBar^{(0)}.
\end{split}
\]
Replacing $T$ with $T^{(0)}$ in $Z_\gen = \frac{\alphaBar^2}{\rhoBar\,\kappaBar}$, we have an one-shot approximation for $Z_\gen$:
\begin{equation}\label{one-shot-Zgen}
Z_\gen \approx \frac{\alphaBar^2}{\rhoBar^{(0)}\,\kappaBar^{(0)}} = \frac{ \left( \int \alpha \,dT \right)^2 }{\Delta T \,\int \rho \kappa \,dT} =: Z_\gen^{(0)}.
\end{equation}

\emph{To approximate $\tau$, we assume} the Seebeck coefficient is a linear function of $T$:
\[ \alpha(T) \approx \alpha_\lin(T) := \alpha_h + \left( \frac{\alpha_c - \alpha_h}{T_c - T_h} \right) \left(T- T_h \right). \]
In this way we can observe the effect of the gradient of $\alpha$ on $\tau$ more clearly. Since the $\tau$ in \eqref{SI-tau} has $K\,\delta T^{(1)}$ term, we estimate a relevant term:
\[
\begin{split}
F_T^{(1)}(s) &\approx \int_0^s \frac{1}{A} T \frac{d\alpha_\lin}{dT}(T(x)) \frac{dT}{dx}\,dx = \int_{T_h}^{T(s)} \frac{1}{A} T \frac{\alpha_c-\alpha_h}{T_c-T_h} \,dT\\
&= \frac{1}{2A} \frac{\alpha_c-\alpha_h}{T_c-T_h} (T(s)^2 - T_h^2) =: \widehat{F^{(1)}}(T(s)).
\end{split}
\]
\emph{Using} $-\kappa \frac{dT}{dx} \approx \kappaBar^{(0)} \frac{\Delta T}{L}$ from \eqref{SI-heat-flux-approx},
\[
\begin{split}
\delta T^{(1)} &= \int_0^L \frac{F_T^{(1)}(x)}{\kappa(x)}\,dx \approx - \int_0^L \frac{\widehat{F^{(1)}}(T(x))}{\kappaBar^{(0)}} \frac{L}{\Delta T}  \frac{dT}{dx}\,dx\\
&= \frac{1}{\kappaBar^{(0)}} \frac{L}{\Delta T} \int_{T_c}^{T_h} \widehat{F^{(1)}}(T)\,dT \\
&= \frac{1}{2 K^{(0)}} \frac{1}{\Delta T} \frac{\alpha_c-\alpha_h}{T_c-T_h} \frac{1}{3} (\Delta T)^2 (-3T_h +\Delta T) \\
&= \frac{\alpha_h-\alpha_c}{6 K^{(0)}} (-3T_h +\Delta T)=: \widehat{\delta T^{(1)}}
\end{split}
\]
where $K^{(0)} := \frac{A}{L}\kappaBar^{(0)}$. Therefore we have an one-shot approximation for $\tau$:
\[
\begin{split}
\tau &\approx \frac{1}{\overline{\alpha_\lin} \Delta T} \left[ (\overline{\alpha_\lin} - \alpha_h) T_h -K^{(0)} \,\widehat{\delta T^{(1)}} \right]\\
&= -\frac{1}{3} \frac{\alpha_h-\alpha_c}{\alpha_h+\alpha_c} =: \tau_{\rm lin}^{(0)}.
\end{split}
\]

\emph{To approximate $\beta$, we assume} the $\rho \kappa$ is a linear function of $T$:
\[ (\rho\kappa)(T) \approx (\rho\kappa)_\lin(T) := (\rho\kappa)_h + \left( \frac{(\rho\kappa)_c - (\rho\kappa)_h}{T_c - T_h} \right) \left(T- T_h \right). \]
\emph{Using} $-\kappa \frac{dT}{dx} \approx \kappaBar^{(0)} \frac{\Delta T}{L}$ from \eqref{SI-heat-flux-approx}, we approximate relevant terms for $\beta$:
\[
\begin{split}
F_T^{(2)}(s) &= \int_0^s \frac{1}{A^2}(\rho\kappa)(T(x)) \frac{1}{\kappa(x)} \,dx \approx \frac{-L}{A^2 \kappaBar^{(0)} \Delta T} \int_0^s (\rho\kappa)_\lin(T(x)) \frac{dT}{dx}\,dx\\
&= \frac{-L}{A^2 \kappaBar^{(0)} \Delta T} \int_{T_h}^{T(s)} (\rho\kappa)_\lin(T) \,dT \\
&= \frac{-L}{A^2 \kappaBar^{(0)} \Delta T} \Big[ (\rho\kappa)_h (T(s)-T_h) + \frac{1}{2} \frac{(\rho\kappa)_c-(\rho\kappa)_h}{T_c-T_h} (T(s)-T_h)^2 \Big]\\
& =: \widehat{F^{(2)}}(T(s))
\end{split}
\]
hence
\[
\begin{split}
\delta T^{(2)} &= \int_0^L \frac{F_T^{(2)}(x)}{\kappa(x)}\,dx \approx \int_0^L \widehat{F^{(2)}}(T(x)) \Big(-\frac{L}{\kappaBar^{(0)} \Delta T} \Big) \frac{dT}{dx}\,dx\\
&= \frac{-L}{\kappaBar^{(0)} \Delta T} \int_{T_h}^{T_c} \widehat{F^{(2)}}(T)\,dT \\
&= \frac{1}{6(K^{(0)})^2} \big( 2(\rho\kappa)_h + (\rho\kappa)_c \big) =: \widehat{\delta T^{(2)}}.
\end{split}
\]
Therefore we have an one-shot approximation for $\beta$:
\[
\begin{split}
\beta &\approx \frac{2}{\frac{L}{A}\rhoBar^{(0)}} K^{(0)} \,\widehat{\delta T^{(2)}} - 1 = \frac{1}{3 \,\rhoBar^{(0)}\kappaBar^{(0)}} (2(\rho\kappa)_h +(\rho\kappa)_c) -1\\
&\approx \frac{1}{\frac{3}{2} ((\rho\kappa)_h+(\rho\kappa)_c)} (2(\rho\kappa)_h +(\rho\kappa)_c) -1\\
&= \frac{1}{3} \frac{(\rho\kappa)_h - (\rho\kappa)_c}{(\rho\kappa)_h +(\rho\kappa)_c} =: \beta_{\rm lin}^{(0)}.
\end{split}
\]
In summary, we have one-shot approximations as following:
\begin{equation}\label{SI-one-shot-approx}
Z_\gen \approx Z_\gen^{(0)} \equiv \frac{ \left( \int \alpha \,dT \right)^2 }{\Delta T \,\int \rho \kappa \,dT}, 
\quad
\tau \approx \tau_{\rm lin}^{(0)} \equiv -\frac{1}{3}\frac{\alpha_h - \alpha_c}{\alpha_h+\alpha_c}, \quad 
\beta \approx \beta_{\rm lin}^{(0)} \equiv \frac{1}{3}\frac{\rho_h\kappa_h-\rho_c\kappa_c}{\rho_h\kappa_h+\rho_c\kappa_c}.
\end{equation}

\begin{figure}
\centering \includegraphics[width=0.8\textwidth]{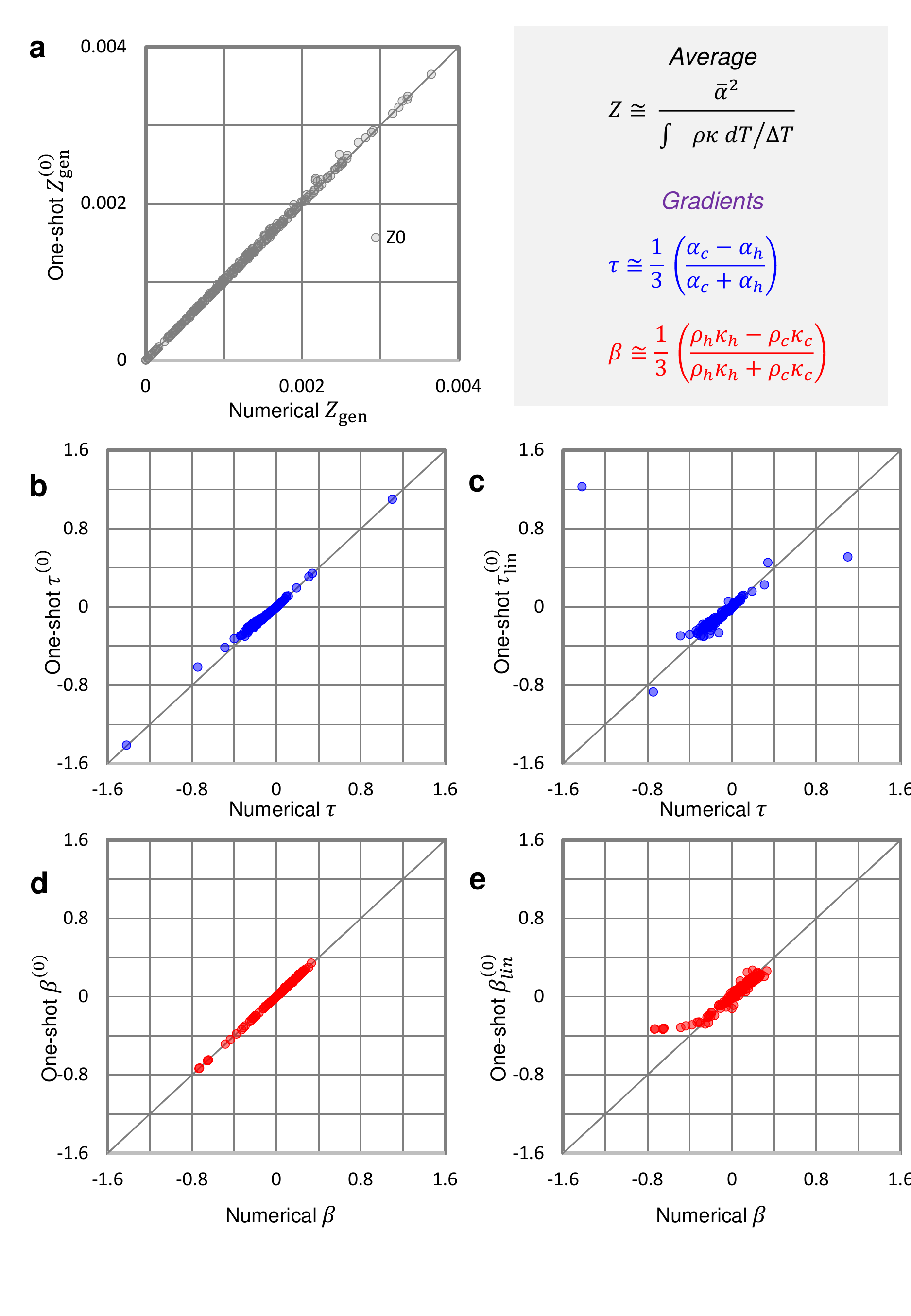}
\caption{Estimation of thermoelectric degrees of freedom for 276 materials. Numerical $Z_\gen$, $\tau$, $\beta$ are computed using the exact $T$ at the maximum efficiency. One-shot approximations $Z_\gen^{(0)}$, $\tau^{(0)}$, $\beta^{(0)}$ are computed using the $T^{(0)}$ for $J=0$. Going further, the $\tau_{\rm lin}^{(0)}$ and $\beta_{\rm lin}^{(0)}$ are computed by assuming the linearity of $\alpha$ and $\rho\kappa$; see \eqref{one-shot-approx} for their explicit formula.
}\label{fig-one-shot-approx}
\end{figure}

The \emph{one-shot approximation} derived above is accurate enough for many cases.
See Figure \ref{fig-one-shot-approx}, where we compare the exact $Z_{\rm gen}$, $\tau$, $\beta$ with their one-shot approximations for 276 materials.

\begin{figure}
\centering \includegraphics[width=\textwidth]{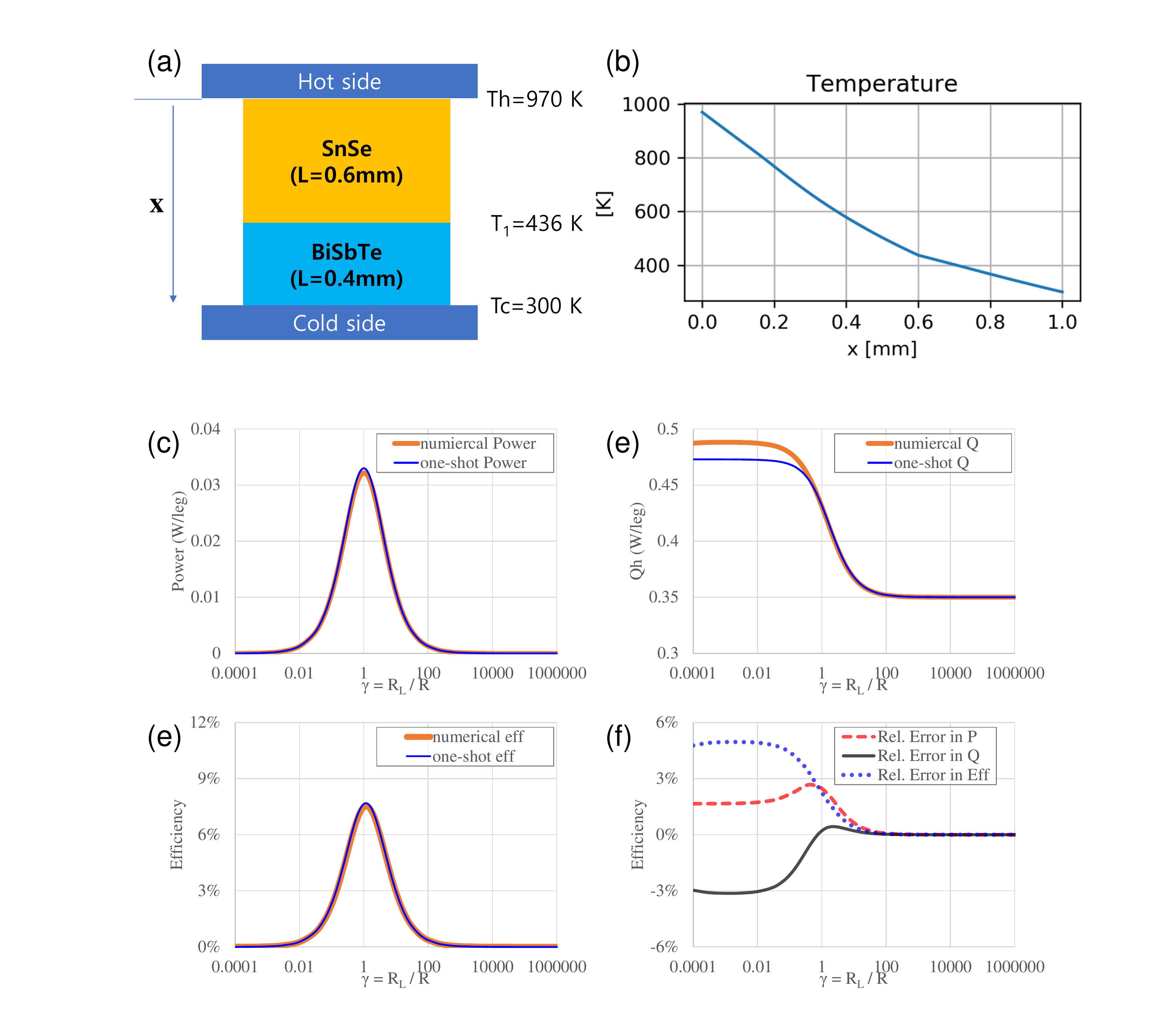}
\caption{
The thermoelectric performances of a two-stage segmented leg predicted by the one-shot approximation. The numerical exact values are computed by fixed-point iteration and the one-shot values are computed using $Z_\gen^{(0)}$, $\tau_{\rm lin}^{(0)}$ and $\beta_{\rm lin}^{(0)}$; see \eqref{one-shot-approx} for the explicit one-shot formula.
(a) The geometry of the segmented leg: $\rm SnSe$ \cite{zhao_ultralow_2014} and $\rm BiSbTe$ \cite{poudel_high-thermoelectric_2008} are used for hot and cold-side materials. $T_h = 970 K$ and $T_c = 300 K$ are used.
(b) Exact temperature distribution obtained by solving the integral equation \eqref{SI-T-integral-form} of $T$ with fixed-point iteration.
(c) Power delivered outside, (d) heat current at the hot side, (e) efficiency, and (f) relative errors in power, heat current, efficiency between the numerical value and the one-shot approximation.
}\label{fig-segQ}
\end{figure}

Furthermore, these one-shot approximations can be used to predict the performance of \emph{segmented} devices.
In Figure \ref{fig-segQ}, we consider a two-stage segmented leg with no contact resistance. The segmented leg consists of SnSe \cite{zhao_ultralow_2014} for hot side and BiSbTe \cite{poudel_high-thermoelectric_2008} for cold side.
The exact temperature distribution $T$ insdie the leg shows a jump of the gradient at $x=0.6$ due to the inhomogeneity of the material; see Figure \ref{fig-segQ}(b).
Despite the nonlinearity of the $T$, the one-shot approximation using $Z_\gen^{(0)}$, $\tau_{\rm lin}^{(0)}$ and $\beta_{\rm lin}^{(0)}$, which does not use the exact $T$, shows high accuracy in prediction of thermoelectric performances; see Figure \ref{fig-segQ}(c)-(f).
The relative error is high near $\gamma=0$, where the reaction term is large due to the large electric current and thereby large Joule heat. For large $\gamma$, the error is negligible. Near the $\gamma=1$, the error is acceptable; the relative error is less than 5\%. The one-shot approximation predicts the maximum efficiency to be 7.68\% while the exact value is 7.53\%.

\clearpage
\section*{Acknowledgement}
\textcolor{blue}{
B.R. and J.C. contributed equallit to this work. The most of the work is already rerpoted in arXiv:1810.11148 , entitled `'Thermoelectric Efficiency has \emph{Three} Degrees of Freedom'' \cite{ryu2018thermoelectric}. 
}




\end{document}